\documentclass[10pt, conference, compsocconf, balance]{IEEEtran}
\newcommand{\iscasubmissionnumber}{51}

\let\labelindent\relax
\usepackage{hadianeh}
\usepackage{macros}
\usepackage{numbers}
\usepackage{names}

\usepackage[subtle]{savetrees}

\hypersetup{%
  unicode=true,%
  pdfstartview={FitH},%
  colorlinks=true,
  urlcolor  = blue,
  colorlinks=true,
  linkcolor=blue,
  citecolor=blue
}

\renewcommand{\footnoterule}{%
  \kern -1pt
  \hrule width 50pt height 0.4pt
  \kern 2pt
}

\usepackage{array}
\newcolumntype{L}[1]{>{\raggedright\let\newline\\\arraybackslash\hspace{0pt}}m{#1}}
\newcolumntype{C}[1]{>{\centering\let\newline\\\arraybackslash\hspace{0pt}}m{#1}}
\newcolumntype{R}[1]{>{\raggedleft\let\newline\\\arraybackslash\hspace{0pt}}m{#1}}

\let\OLDthebibliography\thebibliography
\renewcommand\thebibliography[1]{
  \OLDthebibliography{#1}
  \setlength{\parskip}{0pt}
  \setlength{\itemsep}{0pt plus 0.3ex}
}
\renewcommand{\IEEEbibitemsep}{0pt plus 0.5pt}
\makeatletter
\IEEEtriggercmd{\reset@font\normalfont\fontsize{8pt}{8.40pt}\selectfont}
\makeatother
\IEEEtriggeratref{1}
\begin{document}

\title{\huge{\vspace{-0.35cm}\protect\taganax:~A Unified MIMD-SIMD Acceleration for\\Generative Adversarial Networks}\vspace{-0.6cm}}

\author{
\begin{tabular}{cccc}
\fontsize{10}{12}\selectfont{}Amir Yazdanbakhsh&\fontsize{10}{12}\selectfont{}Kambiz Samadi$^\ddagger$&\fontsize{10}{12}\selectfont{}Nam Sung Kim$^\flat$&\fontsize{10}{12}\selectfont{}Hadi Esmaeilzadeh$^\S$
\\
\multicolumn{4}{c}{\fontsize{10}{12}\selectfont{}\textbf{A}lternative \textbf{C}omputing \textbf{T}echnologies ({\color[HTML]{0B6121}{\textbf{ACT}}}) Lab}
\end{tabular}
\cr
\begin{tabular}{cccc}
\fontsize{9}{16}\selectfont{}Georgia Institute of Technology&\fontsize{9}{16}\selectfont{}$^\ddagger$Qualcomm Technologies, Inc.&\fontsize{9}{16}\selectfont{}$^\flat$University of Illinois at Urbana-Champaign&\fontsize{9}{16}\selectfont{}$^\S$UC San Diego
\end{tabular}
\IEEEauthorblockN{\vspace{0.1cm}}
\begin{tabular}{cccc}
\cemail{\href{mailto:a.yazdanbakhsh@gatech.edu}{a.yazdanbakhsh@gatech.edu}}
&\cemail{\href{mailto:ksamadi@qti.qualcomm.com}{ksamadi@qti.qualcomm.com}}&\cemail{\href{mailto:nskim@illinois.edu}{nskim@illinois.edu}}&\cemail{\href{mailto:hadi@eng.ucsd.edu}{hadi@eng.ucsd.edu}}	
\end{tabular}
\vspace{-0.4cm}
}

\date{}

\fancypagestyle{firstpage}{
  \fancyhf{}	
   \chead{\it Appears in the Proceedings of the 45$^{th}$ International Symposium on Computer Architecture (ISCA), 2018}
}
\maketitle
\thispagestyle{firstpage}
\pagestyle{plain}
\begin{abstract}
\label{sec:abstract}
Generative Adversarial Networks (GANs) are one of the most recent deep learning models that generate synthetic data from limited genuine datasets.
GANs are on the frontier as further extension of deep learning into many domains (e.g., medicine, robotics, content synthesis) requires massive sets of labeled data that is generally either unavailable or prohibitively costly to collect.
Although GANs are gaining prominence in various fields, there are no accelerators for these new models.
In fact, GANs leverage a new operator, called transposed convolution, that exposes unique challenges for hardware acceleration.
This operator first inserts zeros within the multidimensional input, then convolves a kernel over this expanded array to add information to the embedded zeros.
Even though there is a convolution stage in this operator, the inserted zeros lead to underutilization of the compute resources when a conventional convolution accelerator is employed.
We propose the \ganax architecture to alleviate the sources of inefficiency associated with the acceleration of GANs using conventional convolution accelerators, making the first GAN accelerator design possible.
We propose a reorganization of the output computations to allocate compute rows with similar patterns of zeros to adjacent processing engines, which also avoids inconsequential multiply-adds on the zeros. 
This compulsory adjacency reclaims data reuse across these neighboring processing engines, which had otherwise diminished due to the inserted zeros.
The reordering breaks the full SIMD execution model, which is prominent in convolution accelerators. 
Therefore, we propose a unified MIMD-SIMD design for \ganax that leverages repeated patterns in the computation to create distinct microprograms that execute concurrently in SIMD mode.
The interleaving of MIMD and SIMD modes is performed at the granularity of single microprogrammed operation.
To amortize the cost of MIMD execution, we propose a decoupling of data access from data processing in \ganax. 
This decoupling leads to a new design that breaks each processing engine to an access micro-engine and an execute micro-engine. 
The proposed architecture extends the concept of access-execute architectures to the finest granularity of computation for each individual operand.
Evaluations with six GAN models shows, on average, 3.6$\times$ speedup and 3.1$\times$ energy savings over \eyeriss without compromising the efficiency of conventional convolution accelerators.
These benefits come with a mere $\approx$7.8$\%$ area increase.
These results suggest that \ganax is an effective initial step that paves the way for accelerating the next generation of deep neural models.
\end{abstract}

\begin{IEEEkeywords}
Generative Adversarial Networks; GAN; Accelerators; Dataflow; SIMD-MIMD; Deep Neural Networks; DNN; Convolution Neural Networks; CNN; Transpoed Convolution; Access-Execute Architecture\vspace{-0.0cm}
\end{IEEEkeywords}
\section{Introduction}
\label{sec:intro}  
Deep Neural Networks (DNNs) have been widely used to deliver unprecedented levels of accuracy in various applications.
However, they rely on the availability of copious amount of labeled training data, which can be costly to obtain as it requires human effort to label.
To address this challenge, a new class of deep networks, called Generative Adversarial Networks (GANs), have been developed with the intention of automatically generating larger and richer datasets from a small initial labeled training dataset.
GANs combine a generative model, which attempts to create synthetic data similar to the original training dataset, with a discriminative model, a conventional DNN that attempts to discern if the data produced by the generative model is synthetic, or belongs to the original training dataset~\cite{gan:goodfellow:nips:2014}.
The generative and discriminative models compete with each other in a minimax situation, resulting in a stronger generator and discriminator.
As such, GANs can create new impressive datasets that are hardly discernible from the original training datasets.
With this power, GANs have gained popularity in numerous domains, such as medicine, where overtly costly human-centric studies need to be conducted to collect relatively small labeled datasets~\cite{medical:image:syn,retinal:image:syn}.
Furthermore, the ability to expand the training datasets has gained considerable popularity in robotics~\cite{gan:robot:nips:2016}, autonomous driving~\cite{sadgan:arXiv:2016}, and media synthesis~\cite{artgan:arxiv:2017,gpgan:arxiv:2017,discogan:arxiv:2017,gan:music_synth:arxiv:2017,gan:midinet:arxiv:2017,3dgan:nips:2016,dcgan:arxiv:2015} as well.

Currently, advances in acceleration for conventional DNNs are breaking the barriers to adoption~\cite{brainwave:2017,tpu:isca:2017,eie:isca:2016,eyeriss:isca:2016,dnnweaver:micro:2016,diannao:isca:2014}.
However, while GANs are set to push the frontiers in deep learning, there is a lack of hardware accelerators that address their computational needs.
This paper sets out to explore this state-of-the-art dimension in deep learning from the hardware acceleration perspective.
Given the abundance of the accelerators for conventional DNNs~\cite{bitfusion:isca:2018,eyeriss:jssc:2017,scnn:isca:2017,tetris:asplos:2017,cnn:loop:fpga:2017,flexflow:hpca:2017,pipelayer:hpca:2017,cambricon:micro:2016,eie:isca:2016,truenorth:arxiv:2016,tabla:hpca:2016,eyeriss:isca:2016,prime:isca:2016,dnnweaver:micro:2016,cnvlutin:isca:2016,blocking:systematic:arxiv:2016,deep-comp:iclr:2016,isaac:isca:2016,shidiannao:isca:2015,FPGADeep:fpga:2015,snnap:hpca:2015,nn:general:pact:2015,ngpu:micro:2015,divergent:taco:2015,anpu:isca:2014,diannao:isca:2014,olivier:isca:2013,npu:micro:2012,neuflow:cvpr:2011}, designing an accelerator for GANs will only be attractive if they pose new challenges in architecture design.
By studying the structure of emerging GAN models~\cite{artgan:arxiv:2017,gpgan:arxiv:2017,discogan:arxiv:2017,gan:music_synth:arxiv:2017,gan:midinet:arxiv:2017,3dgan:nips:2016,dcgan:arxiv:2015}, we observe that they use a fundamentally different type of mathematical operator in their generative model, called \emph{transpose convolution}, that operates on multidimensional input feature maps.

The transposed convolution operator aims to extrapolate information from input feature maps, in contrast to the conventional convolution operator which aims to interpolate the most relevant information from input feature maps.
As such, the transposed convolution operator first inserts zeros within multidimensional input feature maps and then convolves a kernel over this expanded input to augment information to the inserted zeros.
The transposed convolution in GANs fundamentally differs from the operators in the backward pass of training conventional DNNs, as these do not insert zeros.
Moreover, although there is a convolution stage in the transposed convolution operator, the inserted zeros lead to underutilization of the compute resources if a conventional convolution accelerator were to be used.
The following highlights the sources of underutilization and outlines the contributions of this paper, making the first accelerator design for GANs.

\begin{figure}
    \centering
    \includegraphics[width=0.49\textwidth]{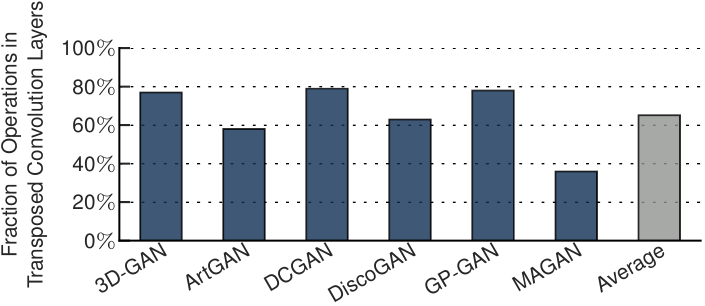}
    \caption{The fraction of multiply-add operations in transposed convolution layers that are inconsequential due to the inserted zeros in the inputs.}
    \vspace{-0.cm}
    \label{fig:zero-data}
\end{figure}

\begin{figure}
    \centering
    \includegraphics[width=0.47\textwidth]{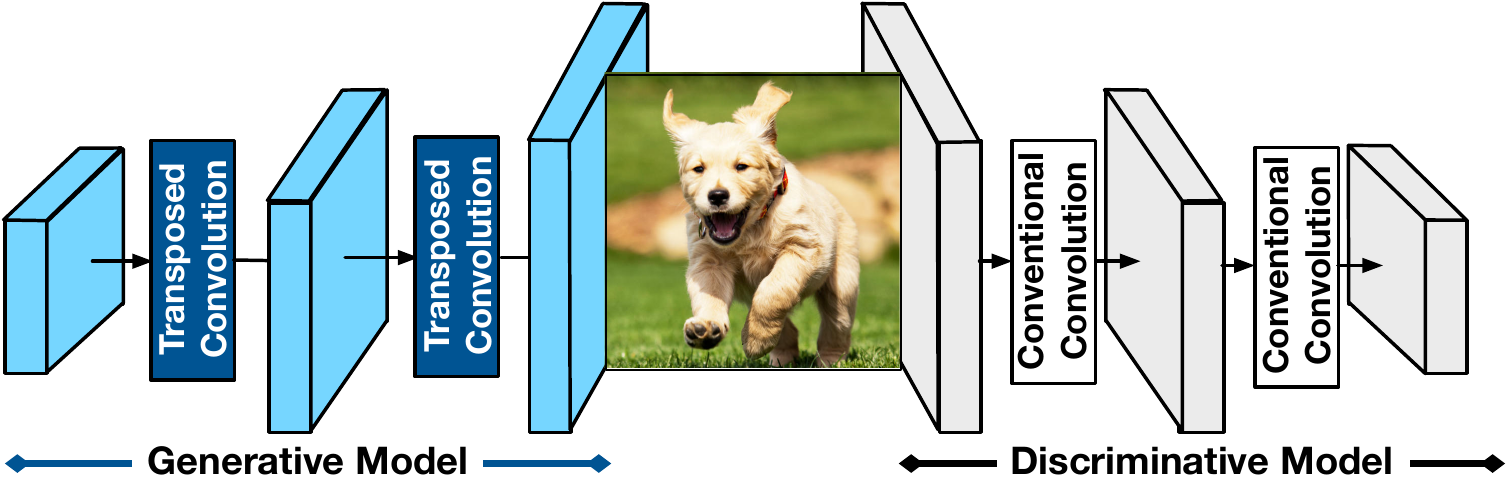}
    \caption{High-level visualization of a Generative Adversarial Network (GAN).}
    \label{fig:dec-gen}
    \vspace{-0.7cm}
\end{figure}

\begin{enumpacked}
	\item \niparagraph{Performing multiply-add on the inserted zeros is inconsequential.} 
	Unlike conventional convolution, the accelerator should skip over the zeros as they constitute more than \xx{60\%} of all the multiply-add operations as Figure~\ref{fig:zero-data} illustrates. Skipping the zeros creates an irregular dataflow and diminishes data reuse if not handled adequately in the microarchitecture. To address this challenge, we propose a reorganization of the output computations that allocates computing rows with similar patterns of zeros to adjacent processing engines. This forced adjacency reclaims data reuse across these neighboring compute units.
	
	\item \niparagraph{Reorganizing the  output computations is inevitable but breaks the SIMD execution model.}
		The inserted zeroes, even with the output computation reorganization, create distinct patterns of computation when sliding the convolution window. As such, the same sequence of operations cannot be repeated across all the processing engines, breaking the full SIMD execution model. Therefore, we propose a unified MIMD-SIMD accelerator architecture that exploits repeated patterns in the computations to create different microprograms that can execute concurrently in SIMD mode. To maximize the benefits from both levels of parallelism, we propose an architecture, called \ganax, that supports interleaving MIMD and SIMD operations at the finest granularity of a single microprogrammed operation. 
	 
	\item \niparagraph{MIMD is inevitable but its overhead needs to be amortized.}
	Changes in the dataflow and the computation order necessitate irregular accesses to multiple different memory structures while the operations are still the same. That is, the data processing part can be SIMD but the irregular data access patterns prevent using this execution model. For \ganax, we propose the decoupling of data accesses from data processing. This decoupling leads to breaking each processing engine into an access micro-engine and an execute micro-engine. The proposed architecture extends the concept of access-execute architecture~\cite{stream:accl:isca:2017,decoupled:affine:isca:2017,decoupled:prefetching:micro:2016,decoupled:smith:sigarch:1982} to the finest granularity of computation for each individual operation.
\end{enumpacked}
Although \ganax addresses these challenges to enable efficient execution of the transposed convolution operator, it does not impose extra overhead, but instead offers the same level of performance and efficiency.
To establish the effectiveness of our architectural innovation, we evaluate \ganax using six recent GAN models, on distinct applications.
On average, \ganax delivers 3.6$\times$ speedup and 3.1$\times$ energy savings over a conventional convolution accelerator.
These results indicate \ganax is an effective step towards designing accelerators for the next generation of deep networks.

\section{Flow of Data in Generative Models}
\label{sec:dataflow}

Generative Adversarial Networks (GANs) have revolutionized modern machine learning by significantly improving generative models while using only limited number of labeled training data.
Figure \ref{fig:dec-gen} shows an overall visualization of a GAN, consisting of two deep neural network models, a generative model and a discriminative model.
These two neural network models oppose each other in a minimax situation.
Specifically, the generative model tries to generate data that will trick the discriminative model to believing the data is from the original training dataset.
Meanwhile, the discriminative model is handed data from either the generative model or the training data and tries to discern between the two.
After these networks compete with each other, they refine their abilities to generate and discriminate, respectively.
This process creates a stronger generative model and discriminative model than could be obtained otherwise~\cite{gan:goodfellow:nips:2014}.
This arrangement of neural networks has opened up many applications, some of which include music generation with accompaniment~\cite{gan:music_synth:arxiv:2017} and the discovery new drugs to cure diseases~\cite{gan:chem:arxiv:2017}.
GANs are enabling our future by pushing forward development in autonomous vehicles, allowing us to imitate human drivers~\cite{gan:auto_imitate:arxiv:2017} and simulate driving scenarios to save testing and training costs~\cite{sadgan:arXiv:2016}.
GANs enable imagination~\cite{gan:imagination:nips:2017}, a major advancement for machine learning and a key step towards true general artificial intelligence.
Here, we overview the challenges and opportunities that were encountered while designing hardware accelerators for GANs.

\niparagraph{Challenges and opportunities for GAN acceleration.}
\begin{figure}
	\captionsetup[subfigure]{justification=centering}
    \centering
    \subfloat[\scriptsize{Conventional Convolution\newline(Data Reduction)}]{\includegraphics[width=0.24\textwidth]{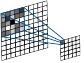}
    \label{fig:conv}}
    \subfloat[\scriptsize{Transposed Convolution\newline(Data Expansion)}]{\includegraphics[width=0.24\textwidth]{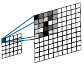}
    \label{fig:tconv}}
    \caption{(a) Convolution operations decreases the size of data (data reduction). (b) Transposed convolution increases the size of data (data expansion).}
    \vspace{-0.5cm}
    \label{fig:conv-tconv}
\end{figure}
%
The generative models in GANs are fundamentally different from the discriminative models.
As Figure~\ref{fig:dec-gen} illustrates, while the discriminative model mostly consists of convolution operations, the generative model uses transposed convolution operations.
Accelerating convolution operations has been the focus of a handful of studies~\cite{bitfusion:isca:2018,eyeriss:jssc:2017,scnn:isca:2017,tetris:asplos:2017,cnn:loop:fpga:2017,flexflow:hpca:2017,pipelayer:hpca:2017,cambricon:micro:2016,eie:isca:2016,truenorth:arxiv:2016,tabla:hpca:2016,eyeriss:isca:2016,prime:isca:2016,dnnweaver:micro:2016,cnvlutin:isca:2016,blocking:systematic:arxiv:2016,deep-comp:iclr:2016,isaac:isca:2016,shidiannao:isca:2015,FPGADeep:fpga:2015,snnap:hpca:2015}; however, accelerating transposed convolution operations has remained unexplored.
Figure~\ref{fig:conv-tconv} depicts the fundamental difference between the conventional convolution and transposed convolution operations.
The convolution operation performs \emph{data reduction} and generally transforms the input data to a smaller representation.
On the other hand, the transposed convolution implements a \emph{data expansion} and transforms the input data to a larger representation.
The transposed convolution operation expands the data by first transforming the input data through inserting zeros between the input rows and columns and then performing the computations by sliding a convolution window over the transformed input data.
Due to this fundamental difference between convolution and transposed convolution operations, using the same conventional convolution dataflow for generative model may lead to inefficiency.
The main reason for such inefficiency can be attributed to the variable number of operations per each convolution window in the transposed convolution.
The variable number of operations per each convolution window is the main result of zero insertion step in transposed convolution.
Because of this zero-insertion step, distinct convolution windows may have a different number of consequential multiplications between inputs and weights.\footnote{A consequential multiplication is a multiplication in which none of the source operands are zero and contributes to the final value of the convolution operation.}
This discrepancy in the number of operations is the root cause for inefficiency in the computations of generative models, if the same convolution dataflow is used.
As such, we aim to design an efficient flow of data for GANs by focusing on: (1) managing the discrepancy in the number of operations per each convolution window in order to mitigate the inefficiencies in the execution of generative models, (2) leveraging the similarities between convolution and transposed convolution operations in order to accelerate both discriminative and generative models on the same hardware platform, and (3) improving the data reuse in discriminative and generative models.

\niparagraph{Why using a conventional convolution dataflow is not efficient for transposed convolution?} 
\begin{figure}
    \centering
    \includegraphics[width=0.49\textwidth]{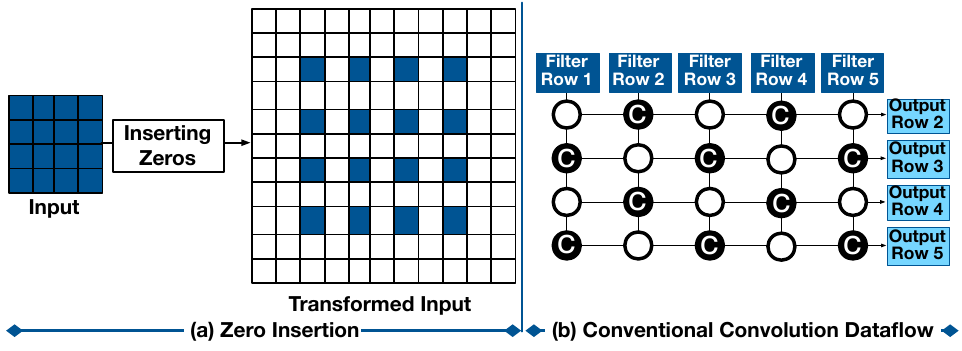}
    \caption{(a) Zero-insertion step in a transposed convolution operation for a 4$\times$4 input and the transformed input. The light-colored squares display zero values in the transformed input. (b) Using conventional dataflow for performing a transposed convolution operation.}
    \vspace{-0.7cm}
    \label{fig:conv_dataflow}
\end{figure}
\begin{figure*}
    \centering
    \includegraphics[width=1.0\textwidth]{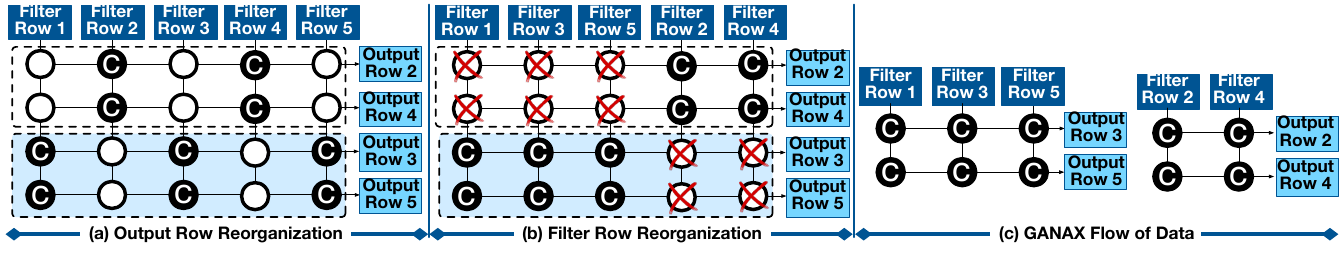}
    \caption{The \ganax flow of data after applying (a) output row reorganization and (b) filter row reorganization. (c) The \ganax flow of data after applying both output and filter row reorganization and eliminating the idle compute nodes. The combination of these flow optimizations reduces the idle (white) compute nodes and improves the resource utilization.}
    \vspace{-0.7cm}
    \label{fig:dataflow}
\end{figure*}
Going through a simple example of a 2-D transposed convolution, we illustrate the main sources of inefficiency in performing transposed convolution, if a conventional convolution dataflow is used.
Figure~\ref{fig:conv_dataflow}(a) illustrates an example of performing a transposed convolution operation using a conventional convolution dataflow.
In this transposed convolution operation, a 5$\times$5 filter with stride of one and padding of two is applied on a 4$\times$4 2D input.
In the initial step, the transposed convolution operation inserts one row and one column of zeros between successive rows and columns (white squares).
Performing this zero-insertion step, the input is expanded from a 4$\times$4 matrix to a 11$\times$11 one.
The number of zeros to be inserted for each transposed convolution layer in the generative models may vary from one layer to another and is a parameter of the network.
After performing the zero-insertion, the next step is to slide a convolution window over the transformed input and perform the multiply-add operations.
Figure~\ref{fig:conv_dataflow}(b) illustrates performing this convolution operation using a conventional convolution dataflow~\cite{eyeriss:jssc:2017,tetris:asplos:2017,eyeriss:isca:2016}.
To avoid clutter in Figure~\ref{fig:conv_dataflow}(b), we only show the dataflow for generating the output rows 2-5.

Each circle in Figure~\ref{fig:conv_dataflow}(b) represents a compute node that can perform vector-vector multiplications between a row of the filter and a row of the zero-inserted input.
The filter rows are spatially reused across each of the computation nodes in a vertical manner.
Once a vector-vector multiplications finish, the partial sums are aggregated horizontally to yield the results of performing transposed convolution operation for each output row.
The black circles represent the compute nodes that are performing consequential operations, whereas the white circles which represent the compute nodes performing inconsequential operations.
As depicted in Figure~\ref{fig:dataflow}(b), there will be inconsequential operation (white circles) if a conventional convolution dataflow is used for the execution of transposed convolution operations.
Because of the inserted zeros, some of the filter rows are not used to compute the value of an output row.
For example, since the 1$^\text{st}$, 3$^\text{rd}$, and 5$^\text{th}$ rows of the input are zero, the 2$^\text{nd}$ output row only needs to perform the operations for non-zero elements; hence using only the 2$^\text{nd}$ and 4$^\text{th}$ filter rows, leaving three compute nodes idle.
Overall, in this example, 50$\%$ of the compute nodes remain idle during the execution of this transposed convolution operation.
Analyzing this transposed convolution operation reveals three main sources of inefficiency when a conventional convolution dataflow is used.

\begin{enumpackedp}
	\item \textbf{Coarse-grain resource underutilization:} Since the consequential filter rows vary from one output row to another, a significant number of compute nodes remain idle. In the aforementioned example, this underutilization applies to 50$\%$ of the compute nodes, which perform vector-vector multiplications.  
	\item \textbf{Fine-grain resource underutilization:} Even within a compute node a large fraction of the multiply-add operations are inconsequential due to the columnar zero insertion.
	\item \textbf{Reuse reduction:} While the compute units pass along the filter rows for data reuse, the inserted zeros render this data transfer futile.
\end{enumpackedp}

\noindent{}
We address the first two sources of inefficiency with a series of optimizations on the flow of data in GANs.
Then, to address the last source of inefficiency that arises because of the inconsequential multiply-add operations within each compute node, we introduce an architectural solution (Section~\ref{sec:arch}).

\niparagraph{Flow of data for generative models in GANAX.}
Figure~\ref{fig:dataflow} illustrates the proposed flow of data optimizations for generative models in \ganax.
To mitigate the challenges of using conventional convolution dataflow for transposed convolution operations in generative models, we leverage the insight that even though the patterns of computation may vary from one output row to another, they are still structured.
Taking a closer look at Figure~\ref{fig:conv_dataflow}, we learn that there are only two distinct patterns\footnote{The location of white and black circles (compute nodes) defines each pattern.} in the output row computations.
In this example, the even output rows (\ie, 2$^\text{nd}$ and 4$^\text{th}$) use one pattern of computation, whereas the odd output rows (\ie, 3$^\text{rd}$ and 5$^\text{th}$) use a different pattern for their computations.
Building upon this observation, we introduce a series of flow of data optimizations to mitigate the aforementioned inefficiencies in the computation of transposed convolution operation, if a conventional convolution dataflow used.

The first optimization maximizes the data reuse by reorganizing the computation of the output rows in a way that the rows with the same pattern in their computations become adjacent.
Figure~\ref{fig:dataflow}(a) illustrates the flow of data after applying this output row reorganization.
Applying the output row reorganization in this example, make the even-indexed (2$^\text{nd}$ and 4$^\text{th}$ output rows) output rows adjacent. Similar adjacency is established for odd-indexed  (3$^\text{rd}$ and 5$^\text{th}$ output rows) output rows.
Although this optimization addresses the data reuse problem, it does not deal with the resource underutilization (\ie, idle compute nodes (white circles) still exist).
To mitigate this resource underutilization, we introduce the second optimization that reorganizes the filter rows.
As shown in Figure~\ref{fig:dataflow}(b), applying the filter row reorganization establishes an adjacency for the 1$^\text{st}$, 3$^\text{rd}$, and 5$^\text{th}$ filter rows. Similarly, the 2$^\text{nd}$ and 4$^\text{th}$ filter rows become adjacent.
After applying output and filter row reorganization, as shown in Figure~\ref{fig:dataflow}(b), the idle compute nodes can be simply eliminated from the dataflow.
Figure~\ref{fig:dataflow}(c) illustrates the \ganax flow of data after performing both optimizations, which improves the resource utilization for transposed convolution operation from 50\% to 100\%.

The proposed \ganax flow of data also addresses the inefficiency in performing the horizontal accumulation of partial sums.
As shown in Figure~\ref{fig:conv_dataflow}(b), the conventional convolution dataflow requires five cycles to perform the horizontal accumulation for each output row, regardless of their locations.
However, comparing Figure~\ref{fig:conv_dataflow}(b) and Figure~\ref{fig:dataflow}(c), we observe that after applying output and filter row reorganization optimizations, the number of required cycles for performing the horizontal accumulation reduces from five to two for even-indexed output rows and from five to three for odd-indexed output rows.
While the proposed flow of data optimizations effectively improve the resource utilization for transposed convolution, there arises an interesting architectural challenge: \emph{how to fully utilize the parallelism between the computations of the output rows that require different number of cycles for horizontal accumulation (two cycles for even-indexed and three cycles for odd-indexed output rows)?}
If a SIMD execution model is used, some of the compute nodes have to remain idle until the accumulations for the output rows that require more cycles for horizontal accumulation, finish.
The next section elaborates on the \ganax architecture that exploits the introduced flow of data for transposed convolution and fully utilize the parallelism between distinct output rows by conjoining the MIMD and SIMD execution models.
\section{Architecture Design for GANAX}
\label{sec:arch}
The execution flow of the generative model (\emph{i.e.} zero-insertion and variable number of operations per each convolution window) in GANs poses unique architectural challenges that the traditional convolution accelerators~\cite{tetris:asplos:2017,eyeriss:jssc:2017,eyeriss:isca:2016,dnnweaver:micro:2016,diannao:isca:2014} can not adequately address.
There are two fundamental architectural challenges for GAN acceleration as follows:

\begin{figure}
    \centering
    \includegraphics[width=0.47\textwidth]{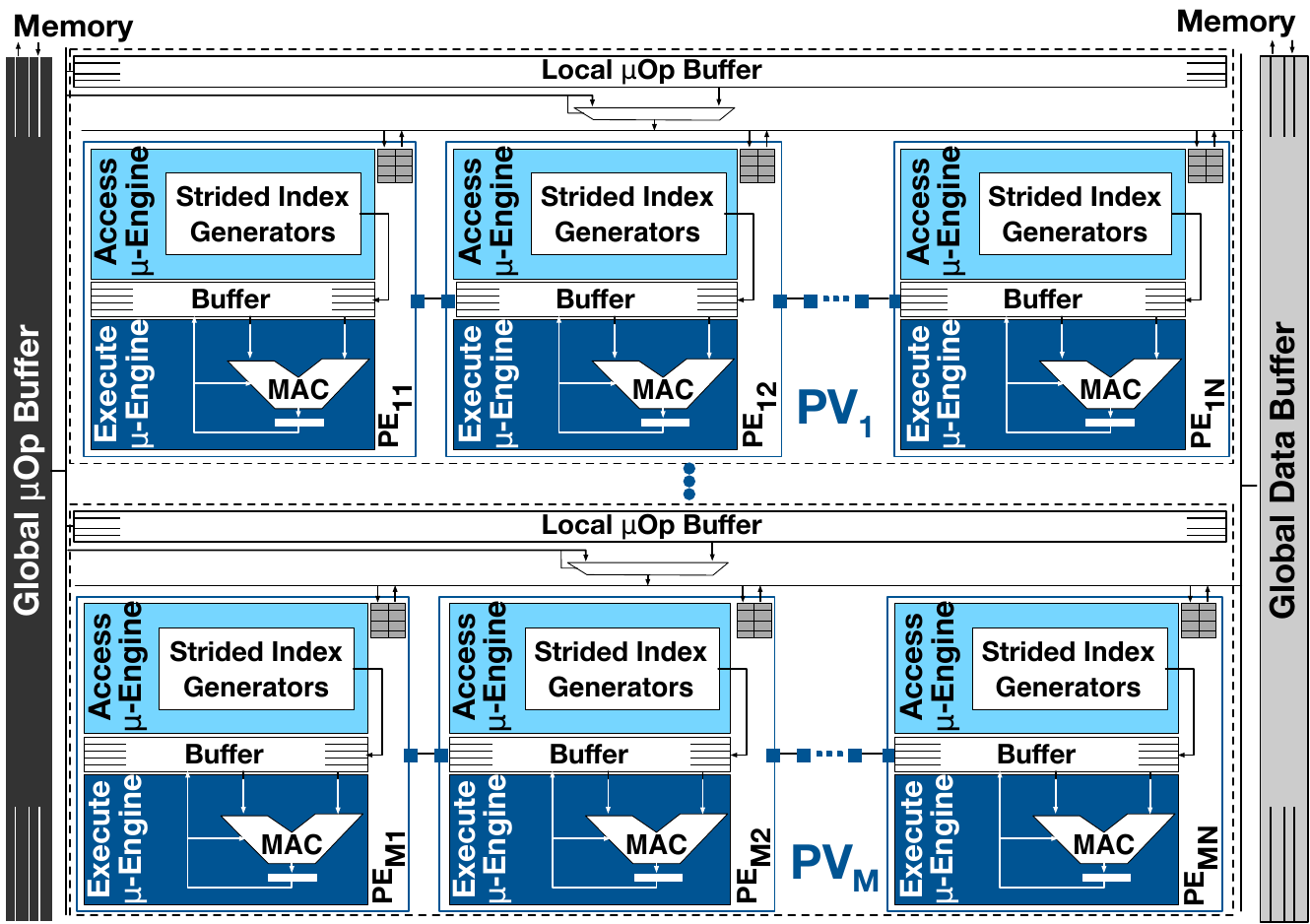}
    \caption{Top-level block diagram of \ganax architecture.}
    \label{fig:tlganax}
    \vspace{-0.8cm}
\end{figure}
\niparagraph{Resource underutilization.}
		The first challenge arises due to the variable number of operations per each convolution window in transposed convolution operation.
		In most of recent accelerators~\cite{eyeriss:jssc:2017,tetris:asplos:2017,dnnweaver:micro:2016,diannao:isca:2014}, which mainly target conventional convolution operation, the processing engines generally work in a SIMD manner.
		The convolution windows in conventional convolution operation follow a regular pattern and the number of operations for each of these windows remains invariable.
		Due to these algorithmic characteristics of conventional convolution operation, a SIMD execution model is an efficient and practical model.  
		However, since the convolution windows in transposed convolution operations exhibit a variable number of operations, a SIMD execution model is not an adequate design choice for these operations.
		While using a SIMD model utilizes the data parallelism between the convolution windows with the same number of operations, its efficiency is limited in exploiting this execution model for the windows with a different number of operations.
		That is, if one uses a convolution accelerator with a SIMD execution model for transposed convolution operations, the processing engines that are performing the operations for a convolution window with fewer number of operations have to remain idle until the operations for other convolution windows finish.
        To address this challenge, we introduce a unified MIMD-SIMD architecture to accelerate the transposed convolution operation without compromising the efficiency of conventional convolution accelerators for convolution operations.
        This unified MIMD-SIMD architecture effectively maximizes the utilization of accelerator compute resources while effectively utilizing the parallelism between the convolution windows with different number of operations.

\niparagraph{Inconsequential computations.}
		The second challenge emanates from the large number of zeros inserted in the multidimensional input feature map for transposed convolution operations.
		Performing MAC operations on these zeros is inconsequential and wastes accelerator resources (See Figure~\ref{fig:zero-data}), if not skipped.
		We address this challenge by leveraging an observation that even though the data access patterns in transposed convolution operations are irregular, they are still structured.
		Furthermore, these structured patterns are repetitive across the execution of transposed convolutional operations. 
		Building upon these observations, the \ganax architecture decouples the operand access and execution. 
		Each processing engine in this architecture consists of a simple access engine that repetitively generates the addresses for operand accesses without interrupting the execute engine.
In the next sections, we examine these architectural challenges in details for GAN acceleration and expound the proposed microarchitectural solutions.

\subsection{Unified MIMD-SIMD Architecture}
In order to mitigate the resource underutilization, we devise a unified SIMD-MIMD architecture that reaps the benefits of SIMD and MIMD execution models at the same time.
That is, while our architecture executes the operations for convolution windows with distinct computation patterns in a MIMD manner, it performs the operations of the convolution windows with the same computation pattern in a SIMD manner.
Figure~\ref{fig:tlganax} illustrates the high-level diagram of the \ganax architecture, which is comprised of a set of identical processing engines (PE).
The PEs are organized in a 2D array and connected through a dedicated network.
Each PE consists of two $\mu$-engines, namely the access $\mu$-engine and the execute $\mu$-engine.
The access $\mu$-engine generates the addresses for source and destination operands, whereas execute $\mu$-engine merely performs simple operations such as multiplication, addition, and multiply-add.
The memory hierarchy is composed of an off-chip memory and two separate on-chip global buffers, one for data and one for $\mu$ops.
These global on-chip buffers are shared across all the PEs.
Each PE operates on one row of filter and one row of input and generates one row of partial sum values.
The partial sum values are further accumulated horizontally across the PEs to generate the final output value. 
Using a SIMD model for transposed convolution operations leads to resource underutilization.
The PEs that perform the computation for convolution windows with fewer number of operations remains idle, wasting computational resources.
The simple solution is to replace the SIMD model with a fully MIMD computing model and utilize the parallelism between the convolution windows with different number of operations.
However, a MIMD execution model requires augmenting each processing engine with a dedicated operation buffer.
While this design resolves the underutilization of resources, it imposes a large area overhead, increasing area consumption by $\approx$3$\times$.
Furthermore, fetching and decoding instructions from each of these dedicated operation buffers significantly increases the von Neumann overhead of instruction fetch and decode.  
To address these challenges, we design the \ganax architecture upon this observation that PEs in the same row perform same operations for a large period of time. 
As such, the proposed architecture leverages this observation and develop a middle ground between a fully SIMD and a fully MIMD execution model.
The goal of designing the \ganax architecture is multi-faceted: (1) improve the PE underutilization by combining MIMD/SIMD model of computation for transposed convolution operations (2) without compromising the efficiency of SIMD model for conventional convolution operations.
Next, we explain the two novel microarchitectural components that enable an efficient MIMD-SIMD accelerator design for GAN acceleration.

\niparagraph{Hierarchical $\mu$op buffers.}
To enable a unified MIMD and SIMD model of execution, we introduce a two-level $\mu$op buffer.
Figure~\ref{fig:tlganax} illustrates the high-level structure of the two-level $\mu$op buffer.
The two-level $\mu$op buffer consists of a global and a local $\mu$op buffer.
The local and global $\mu$op buffers work cooperatively to perform the computations for GANs.
Each horizontal group of PEs, called processing vector (PV), shares a local $\mu$op buffer, whereas, the global $\mu$op buffer that is shared across all the PVs.
The \ganax accelerator can operate in two distinct modes: SIMD mode and MIMD-SIMD mode.
Since all the convolution windows in the convolution operation have the same number of multiply-adds, the SIMD execution model is a best fit.
As such for this case, the global $\mu$op buffer bypasses the local $\mu$ops and broadcasts the fetched $\mu$op to all the PEs.
On the other hand, since the number of operations varies from one convolution window to another in transposed convolution operation, the accelerator works in MIMD-SIMD mode.
In this mode, the global $\mu$op buffer sends distinct indices to each local $\mu$op buffer.
Upon receiving the index, each local $\mu$op buffer broadcasts a $\mu$op, at the location pointed by the received index, to all the underlying PEs.
Using MIMD-SIMD mode enables the \ganax accelerator to not only utilize the parallelism between the convolution windows with the same number of operations, but also utilize the parallelism across the windows with distinct number of operations.

\niparagraph{Global $\mu$op buffer.}
Before starting the computations of a layer, a sequence of high-level instructions, which defines the structure of each GAN layer, are statically translated into a series of $\mu$ops.
These $\mu$ops are pre-loaded into the global $\mu$op buffer, and then the execution starts.
Each of the $\mu$ops either performs an operation across all the PEs (SIMD) or initiates an $\mu$op in each PV (MIMD-SIMD).  
The initiated operation in the MIMD-SIMD mode may vary from one PV to another.
The SIMD and MIMD $\mu$ops can be stored in the global $\mu$op buffer in any order.
A 1-bit field in the global $\mu$op identifies the type of $\mu$op: SIMD or MIMD-SIMD. 
In the SIMD mode\emdash{}all the PEs share the same $\mu$op globally but execute it on distinct data\emdash the global $\mu$op defines the intended operation to be performed by all the PEs.
In this mode, the local $\mu$op buffer is bypassed and the global $\mu$op are broadcasted to all the PEs at the same time.
Upon receiving the $\mu$op, all the PEs perform the same operation, but on distinct data.
In the MIMD-SIMD mode\emdash{}all the PEs within the same PV share the same $\mu$op but different PVs may execute different $\mu$ops\emdash the global $\mu$op is partitioned into multiple fields (one filed per each PV), each of which defines an index for accessing an entry in the local $\mu$op buffer.
Upon receiving the index, each local $\mu$op buffer retrieves the corresponding $\mu$op stored at the given index and broadcasts it to all the PEs which it controls.
The global $\mu$op buffer is double-buffered so that the next set of $\mu$ops for performing the computations of GAN layer$_{i+1}$ can be loaded into the buffer while the $\mu$ops for GAN layer$_i$ are being executed.

\niparagraph{Local $\mu$op buffer.}
In the \ganax architecture, each PV has a dedicated local $\mu$op buffer.
In the SIMD mode, the local $\mu$op buffers are completely bypassed and all the PEs perform the same operation that are sent from global $\mu$op buffer.
In the MIMD-SIMD mode, each local $\mu$op buffer is accessed at the location specified by a dedicated field in the global $\mu$op.
This location may vary from one local $\mu$op buffer to another. 
Then, the fetched $\mu$op is broadcasted to all the PEs within a PV to perform the same operation but on distinct data.
Each GAN layer may require a distinct sequence of $\mu$ops both globally and locally.
Furthermore, each PE may need to access millions of operands at different locations to perform the computations of a GAN layer.
Therefore, we may need not to only add large $\mu$op buffers to each PE, but also drain and refill the $\mu$op buffers multiple times.
Adding large buffers to the PEs adds a large area overhead, which could have been utilized to improve the computing power of the accelerator.
Also, the process of draining and refilling the $\mu$op buffers imposes a significant overhead in terms of both performance and energy.
To mitigate these overheads, we introduce decoupled access-execute microarchitecture that enables us to significantly reduce the size of $\mu$op buffers and eliminate the need to drain and refill the local $\mu$op buffers for each GAN layer. 
\subsection{Decoupled Access-Execute $\mu$Engines}
\begin{figure}
    \centering
    \includegraphics[width=0.48\textwidth]{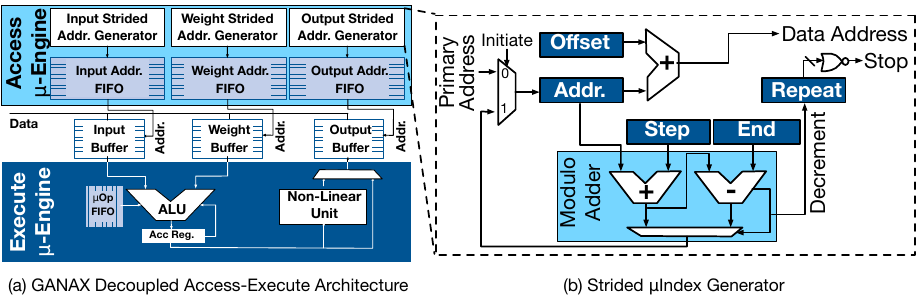}
    \caption{Organization of decoupled Access-Execute architecture.}
    \label{fig:access:execute}
    \vspace{-0.8cm}
\end{figure}

Though the data access patterns in transposed convolution operation are irregular they are still structured.
Furthermore, the data access patterns are repetitive across the convolution windows. 
Building upon this observation, we devise a microarchitecture that decouples the data accesses from from the data processing.
Figure~\ref{fig:access:execute} illustrates the organization of the proposed decoupled access-execute architecture.
The \ganax decoupled access-execute architecture consists of two major microarchitectural units, one for address generation (access $\mu$-engine) and one for performing the operations (execute $\mu$-engine).

The access $\mu$-engine generates the addresses for the input, weight, and output buffers.
The input, weight, and output buffers consume the generated addresses for each data read/write. 
The execute $\mu$-engine, on the other hand, receives the data from the input and weight buffers, performs an operation, and stores the result in the output buffer.
The $\mu$ops of these two engines are entirely segregated.
However, the access and execute $\mu$-engines work cooperatively to perform an operation.
The $\mu$ops for access $\mu$-engine handle the configuration of index generator units.
The $\mu$ops for execute $\mu$engine \emph{only} specify the type of operation to be performed on data.
As such, the execute $\mu$ops do \emph{not} need to include any fields for specifying the source/destination operands.
Every cycle, the access $\mu$engine sends out the addresses for source and destination operands based on its preconfigured parameters.
Then, the execute $\mu$engine performs an operation on the source operands. 
The result of the operation is, then, stored in the location that is defined by the access $\mu$engine.
Having decoupled $\mu$-engines for accessing the data and executing the operations has a paramount benefit of reusing execute $\mu$ops.
Since there is no address field in the execute $\mu$ops, we can reuse the same execute $\mu$op on distinct data over and over again without the need to change any fields in the $\mu$ops.
Reusing the same $\mu$op on distinct data helps to significantly reduce the size of $\mu$op buffers. 

\niparagraph{Access $\mu$-engine.}
Figure~\ref{fig:access:execute} illustrates the microarchitectural units of access $\mu$-engine.
The main function of access $\mu$-engine is to generate the addresses for source and destination operands based on a preloaded configuration.
While designing a full-fledged access $\mu$-engine that is capable of generating various patterns of data addresses enables flexibility for the \ganax accelerator, but it is an overkill for our target application (\ie, GANs).
As mentioned in the dataflow section (Section~\ref{sec:dataflow}), the data access patterns for transposed convolution operations are irregular, yet structured.
Based on our analysis over the evaluated GANs, we observe that the data accesses in the \ganax dataflow are either \emph{strided} or \emph{sequential}.
The stride value for a strided data access pattern depends on the number of inserted zeros in the multidimensional input activation.
Furthermore, these data access patterns are repetitive across a large number of convolution windows and for large number of cycles.  
We leverage these observations to simplify the design of the access $\mu$-engine.
Figure~\ref{fig:access:execute}(a) depicts the block diagram of the access $\mu$engine in \ganax.
The access engine mainly consists of one or more strided $\mu$index generators.
The $\mu$index generator can generate one address every cycle, following a pattern governed by a preloaded configuration.
Since the data access patterns may vary from one layer to another, we design a reconfigurable $\mu$index generator.

Figure~\ref{fig:access:execute}(b) depicts the block diagram of the proposed reconfigurable $\mu$index generator. 
There are five configuration registers that govern the pattern for data address generation.

The \xx{Addr.} configuration register specifies the initial address from which the data address generation starts, while the \xx{Offset} configuration register can be used to offset the range of generated addresses as needed.
The \xx{Step} configuration register specifies the step size between two consecutive addresses, while the \xx{End} configuration register specifies the final value up to which the addresses should be generated.
Finally, the \xx{Repeat} configuration register indicates the number of times that a configured data access pattern should be replayed. 
The modulo adder, which consists of an adder and a subtractor, is used to enable data address generation in a rotating manner.
The modulo adder performs a modulo addition on the values stored in the \xx{Addr.} and \xx{Step} registers.
If the result of this modulo addition is fewer than the value in \xx{End} register, the calculated result is sent to the output. 
This means that the next address to be generated is still within the range of \xx{Addr.} and \xx{End} register values. 
Otherwise, the result of the modulo addition minus the value of \xx{End} register is sent to the output.
That is, the next address to be generated is beyond the \xx{End} register value and the address generation process must start over from the beginning.
In this scenario, the \xx{Decrement} signal is also asserted which cause the value of the \xx{Repeat} register to be decreased by one, indicated one round of address generation is finished.
Once the \xx{Repeat} register reaches zero, the \xx{Stop} signal is asserted and no more addresses are generated.
After configuring the parameters, the strided $\mu$index generator can yield one address per cycle without any further interventions from the controller.
Using this configurable $\mu$index generator along the observation that the data address patterns in GANs are structured, the \ganax architecture can bypass the inconsequential computations and save both cycles and energy.

\niparagraph{Execute $\mu$-engine.}
Figure~\ref{fig:access:execute}(b) depicts the microarchitectural units of execute $\mu$-engine.
The execute $\mu$-engine consists of an ALU, which can perform simple operations such as addition, multiplication, comparison, and multiply-add.
The main job of execute $\mu$-engine is \emph{just} to perform an operation on the received data.
At each cycle the execute $\mu$-engine consumes one $\mu$op from the $\mu$op FIFO and performs the operation on the source operands and store the result back into the destination operand.
If the $\mu$Op FIFO becomes empty, the execute $\mu$op halts and no further operation is performed.
In this case, all the input/weight/output buffers are notified to stop their reads/writes. 
The decoupling between access and execute $\mu$engines enables us to remove the address field from the execute $\mu$ops.
Removing the address field from the execute $\mu$ops allow us to reuse the same $\mu$ops over and over again on different data.
Furthermore, we leverage this $\mu$op reuse and the fact that the computation of the CNN requires a small set of $\mu$ops ($\approx$ 16) to simplify the design of the $\mu$op buffers.
Instead of draining and refilling the $\mu$op buffers, we preload all the necessary $\mu$ops for convolution and transposed convolution operations in the $\mu$op buffers.
For the local $\mu$op buffer, we load \emph{all} the $\mu$ops before starting the computation of a GAN.

\niparagraph{Synchronization between $\mu$engines.}
In the \ganax architecture (Figure~\ref{fig:access:execute}), there is one address FIFO for each strided $\mu$index generator.
The address FIFOs perform the synchronization between access $\mu$-engine and execute $\mu$-engine.
Once an address is generated by a strided $\mu$index generator, the generated address is pushed into the corresponding address FIFO.
The addresses in the address FIFOs are later consumed to read/write data from/into the data buffers (\ie, input/weight/output buffers).
If any of the address FIFOs are full, the corresponding strided $\mu$index generator stops generating new addresses.
In the case that any of the address FIFOs are empty, no data is read/written from/into its corresponding address FIFO.
\section{Instruction Set Architecture Design ($\mu$Ops)}
\label{sec:isa}
The \ganax ISA should provide a set of $\mu$ops to efficiently map the proposed flow of data for both generative and discriminative models onto the accelerator.
Furthermore, these $\mu$ops should be sufficiently \emph{flexible} to serve distinct patterns in the computation for both convolution and transposed convolution operations.
Finally, to keep the size of $\mu$op buffers modest, the set of $\mu$ops should be \emph{succinct}.
To achieve these multifaceted goals, we first introduce a set of algorithmic observations that are associated with GAN models.
Then, we introduce the major $\mu$ops that enable the execution of GAN models on \ganax. 
\subsection{Algorithmic Observations}
The following elaborates a set of algorithmic observations that are the foundation of the \ganax $\mu$ops.

\niparagraph{(1) MIMD/SIMD execution model.}
Due to the regular and structured patterns in the computation across the convolution windows in conventional DNNs, they are best suited for SIMD processing.
However, the patterns in the computation of GANs are inherently different between generative and discriminative models.
Due to the inserted zeros in the generative models, their patterns in the computation vary from one convolutional window to another.
We observe that exploiting a combination of SIMD and MIMD execution model can be more efficient in accelerating GAN models than solely relying on SIMD.      
Therefore, the focus of the \ganax $\mu$ops is to include the operations that enable \ganax to fully utilize the SIMD and MIMD execution models.

\niparagraph{(2) Repetitive computation patterns.}
We observe that even though GANs require a large number of computations, most of these computations are similar between generative and discriminative models.
In addition, these computations are repetitive over a long period of time.
Building upon this observation, we introduce a customized \codebold{repeat} $\mu$op that significant reduces the $\mu$op footprints.
In addition, the commonality between the operations in generative and discriminative models allows us to design a succinct, yet representative, set of $\mu$ops.
To further reduce the $\mu$op footprints, we introduce a dedicated set of execute $\mu$ops that only define the type of operations.
These $\mu$ops are reused for distinct data during the execution of generative and discriminative models on the GANAX architecture.

\niparagraph{(3) Structured and repetitive memory access patterns.}
We observe that despite the irregularity of memory access patterns in generative models, they are still structured and repetitive.
Analyzing the data access patterns of various GANs reveals that their memory access patterns are either sequential or strided.
Building upon this observation and our decoupled access-execute architecture, we introduce a set of access $\mu$ops that are used merely to configure the access $\mu$engines and initiate the address generation process. 
Once initiated, the access $\mu$engines generate the configured access patterns over and over until they are intervened.

\subsection{Access $\mu$Ops}
\ganax access $\mu$ops are used to configure the access $\mu$engine and initiate/stop the process of address generation.
These $\mu$ops are executed across all the PEs within a PV whose index is indicated by \code{pv\_index} field in the $\mu$ops.
Furthermore, in all of these $\mu$ops, \code{\%addrgen\_idx} specifies the index of the targeted address generator in the access $\mu$engine.
The supported $\mu$ops in the access $\mu$engines are as follows:
\begin{enumpacked}
\item \codebold{access.cfg} \code{\%pv\_idx, \code{\%addrgen\_idx,} \code{\%dst,} imm}: This $\mu$op loads a 16-bit \code{imm} value into one of the five \code{\%dst} configuration registers (\emph{i.e.,} as shown in Figure~\ref{fig:access:execute}(b), these configuration registers are \xx{Addr.}, \xx{Offset}, \xx{Step}, \xx{End}, and \xx{Repeat}) of one of the address generators in the access $\mu$engine. 
\item \codebold{access.start} \code{\%pv\_idx,} \code{\%addrgen\_idx}: This $\mu$op initiates the address generation in one of the address generators in the access $\mu$engine. The process of address generation continues until an \codebold{acceess.stop} $\mu$op is executed or the iteration register reaches zero.
\item \codebold{access.stop} \code{\%pv\_idx,} \code{\%addrgen\_idx}: This $\mu$op intervenes the address generation of one of the address generators in the access $\mu$engine. The address generation can be re-initiated again by executing an \code{access.start} $\mu$op.
\end{enumpacked}
\subsection{Execute $\mu$Ops}
Execute $\mu$ops are categorized into two groups:
(1) SIMD $\mu$ops are fetched from each PE's local $\mu$op buffer and executed locally within each PE and (2) the MIMD $\mu$ops are fetched from the global $\mu$op buffer and executed across all PEs.  
The SIMD $\mu$ops can be executed in the MIMD manner as well.
That is, the MIMD $\mu$ops are a superset of the SIMD $\mu$ops.
We first introduce the SIMD $\mu$ops, then explain the extra $\mu$ops that belong to the MIMD group.

\niparagraph{SIMD $\mu$ops.}
SIMD group only comprises a succinct, yet representative set of $\mu$ops for performing convolution and transposed convolution operations.
The combination of SIMD $\mu$ops and the decoupled access-execute architecture in \ganax helps to reduce the size of local $\mu$op buffers.  
The SIMD $\mu$ops do not have source or destination fields and only specify the type of operation to be executed.
Once executed, depending on the type of operation, a given PE consumes the generated addresses by the $\mu$index generators and delivers the data to the execute $\mu$engine.
Since these $\mu$ops do not have any source or destination register, they are pre-loaded into the local $\mu$op buffers before starting the execution.
Then, they are re-used over and over, on distinct data whose addresses are generated by the access $\mu$engines.
The SIMD $\mu$ops are as follows:
\begin{enumpacked}
\item \codebold{add}, \codebold{mul}, \codebold{mac}, \codebold{pool}, and \codebold{act}: Depending on the type, these $\mu$ops consume one or more addresses from the $\mu$index generators for source and destination operands. For example, \codebold{add} consumes two addresses for the source operands and one address for the destination operand, but \codebold{act} uses one address for the source operand and one address for the destination operand.
\item \codebold{repeat}: This $\mu$op causes the next fetched $\mu$op to be repeated a specified number of times. This number is specified in a microarchitectural register in each PE. This register is pre-loaded with a MIMD $\mu$op before the execution starts.  
\end{enumpacked}
\niparagraph{MIMD $\mu$ops.}
The MIMD $\mu$ops are loaded into the global $\mu$op buffers and executed globally across all the PEs.
In addition to all the SIMD $\mu$ops, the following $\mu$ops execute in a MIMD manner: 
\begin{enumpacked}
\item \codebold{mimd.ld} \code{\%pv\_idx,} \code{\%dst, imm}: This $\mu$op loads the  immediate value (\code{imm}) into one of the microarchitectural registers (\code{\%dst}) of all the PEs with a PV. The \code{\%pv\_idx,} specifies the index of the target PV.
This $\mu$op is mainly used to load an immediate value into the repeat register.
\item \codebold{mimd.exe} \code{\%$\mu$op\_index\textsubscript{1},..., \code{\%$\mu$op\_index\textsubscript{i}}}:
Upon receiving this $\mu$op, the i\textsuperscript{th} PV fetches a $\mu$op located at location \code{\%$\mu$op\_index\textsubscript{i}} from its local $\mu$op buffer and executes it across all the PEs horizontally.
Since the value of the \code{\%$\mu$op\_index} may vary from one PV to another, this $\mu$op causes \ganax to operate in a MIMD manner. 
\end{enumpacked}

\begin{table}
    \centering
    \caption{The evaluated GAN models, their released year, and the number of convolution (\code{Conv}) and transposed convolution (\code{TConv}) layers per generative and discriminative models.}
	\vspace{-0.2cm}
    \includegraphics[width=0.48\textwidth]{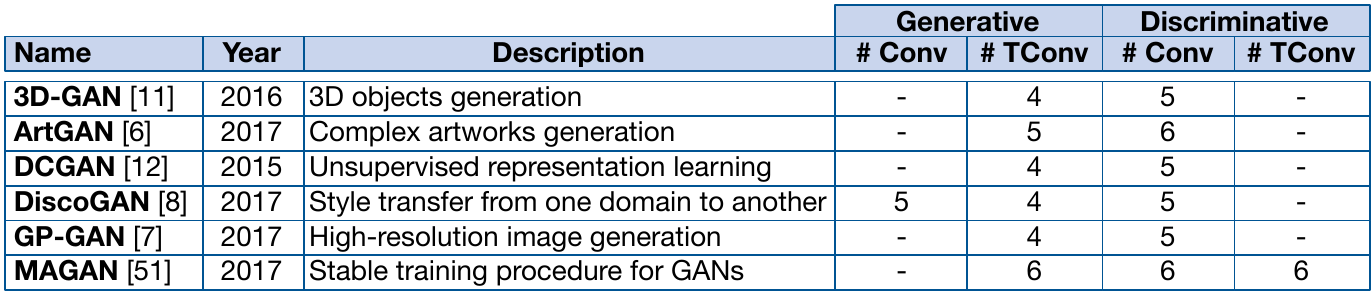}
    \label{tab:bench}
    \vspace{-0.65cm}
\end{table}
\section{Methodology}
\label{sec:eval}
\niparagraph{Workloads.}
We use several state-of-the-art GANs to evaluate the \ganax architecture.
Table~\ref{tab:bench}, shows the evaluated GANs, a brief description of their applications, and the number of convolution (\code{Conv}) and transposed convolution (\code{TConv}) layers per generative and discriminative models.

\niparagraph{Hardware design and synthesis.}
We implement the \ganax microarchitectural units including the strided $\mu$index generator, the arithmetic logic of the PEs, controllers, non-linear function, and other logic hardware units in Verilog.
We use \code{TSMC 45\,nm} standard-cell library and \code{Synopsys Design Compiler (L-2016.03-SP5)} to synthesize these units and obtain the area, delay, and energy numbers.

\niparagraph{Energy measurements.}
Table~\ref{tab:energy} shows the energy numbers for major micro-architectural units, memory operations, and buffer accesses in \code{TSMC 45nm} technology.
To measure the area and read/write access energy of the register files, SRAMs, and local/global buffers, we use \bench{CACTI-P}~\cite{cactip}.
To have a fair comparison, we use energy numbers reported in \tetris~\cite{tetris:asplos:2017}, which has a similar PE architecture as \eyeriss.
In Table~\ref{tab:energy}, the energy overhead of strided $\mu$index generators is included in the normalized energy cost of PE.
For DRAM accesses, we use the Micron's DDR4 system power calculator~\cite{micron:ddr4}.
The same frequency (\xx{500 MHz}) is used for both \eyeriss and \ganax in all the experiments.
\begin{figure}[t]
    \centering
    \subfloat[Speedup]{\includegraphics[width=0.48\textwidth]{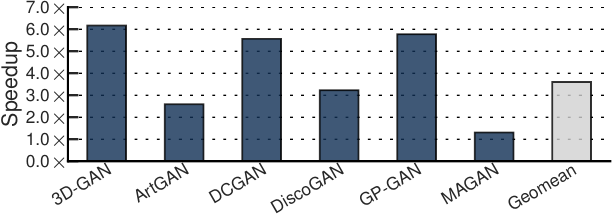}
    \label{fig:speedup-eyeriss}}
    \vspace{-0.0cm}
    \\
     \subfloat[Energy Reduction]{
    \includegraphics[width=0.48\textwidth]{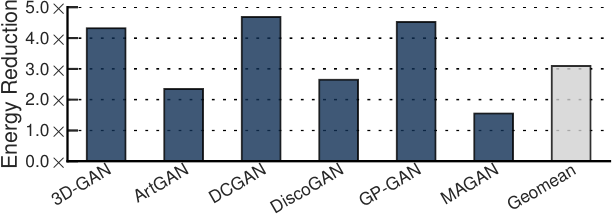}
    \label{fig:energy-eyeriss}}
    \caption{Speedup and energy reduction of generative models compared to \eyeriss~\cite{eyeriss:isca:2016}.}
    \label{fig:speedup-energy-eyeriss}
    \vspace{-0.5cm}
\end{figure}
\niparagraph{Architecture configurations.}
In this paper, we study a configuration of \ganax with 16 Processing Vectors (PVs) each with 16 Processing Engines (PEs).
We use the default \eyeriss configurations for on-chip memories such as the size of input and partial sum registers, weight SRAM, and global data buffer.
The same on-chip memory sizes are used for \ganax.
Each local $\mu$op buffer has 16 entries.
The number of entries is sufficient to encompass all the execute $\mu$ops.
The global $\mu$op buffer has 32 entries each with 64 bits, four bits per each PV.
Each local $\mu$op uses these four bits to index its local $\mu$op buffer.
An extra one bit in the global $\mu$ops determines the execution model of the accelerator for the current operation (\ie, SIMD or MIMD-SIMD).

\niparagraph{Area analysis.}
Table~\ref{tab:area} shows the major architectural components for the baseline architecture (\eyeriss~\cite{eyeriss:jssc:2017,eyeriss:isca:2016}) and \ganax in \code{45\,nm} technology node.
For logic of the microarchitectural units, we use the reported area from the synthesis.
For the memory elements, we use \code{CACTI-P}~\cite{cactip} and the reported numbers in \eyeriss~\cite{eyeriss:jssc:2017}.
In order to be consistent in the results, we scaled down the reported area numbers in \eyeriss from \code{65\,nm} to \code{45\,nm}.
To have a fair comparison between \eyeriss and \ganax, the same number of PEs and on-chip memory are used for both accelerators.
Under this setting, \ganax has an area overhead of \ganaxarea compared to \eyeriss.

\niparagraph{Microarchitectural simulation.}
Table~\ref{tab:area} shows the major microarchictural parameters of \ganax.
We implement a microarchitectural simulator on top of the \eyeriss simulator~\cite{tetris:asplos:2017}.
The extracted energy numbers from logic synthesis and \bench{CACTI-P} are integrated into the simulator to measure the energy consumption of the evaluated network models on \ganax.
To evaluate our proposed accelerator, we extend the \eyeriss simulator with the proposed ISA extensions and the \ganax flow of data.
For all the baseline numbers, we use the plain version of the simulator. 

\label{sec:eval}

\begin{table}
    \centering
    \caption{Energy comparison between \ganax microarchitectural units and memory. PE energy includes the energy consumption of an arithmetic operation and the strided $\mu$index generators.}
    \includegraphics[width=0.49\textwidth]{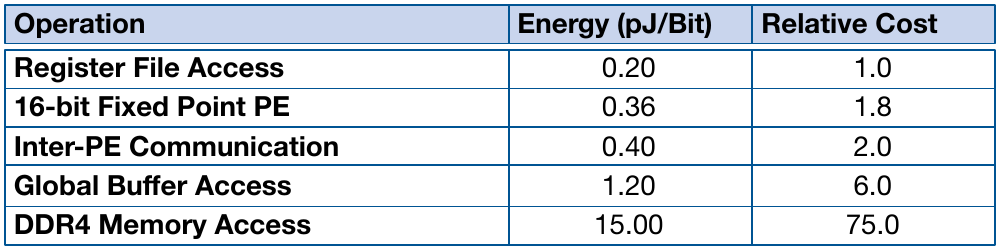}
    \vspace{-0.5cm}
    \label{tab:energy}
\end{table}

\begin{table}
    \centering
    \caption{Area measurement of the major hardware units with \code{TSMC\,45nm}.}
    \includegraphics[width=0.49\textwidth]{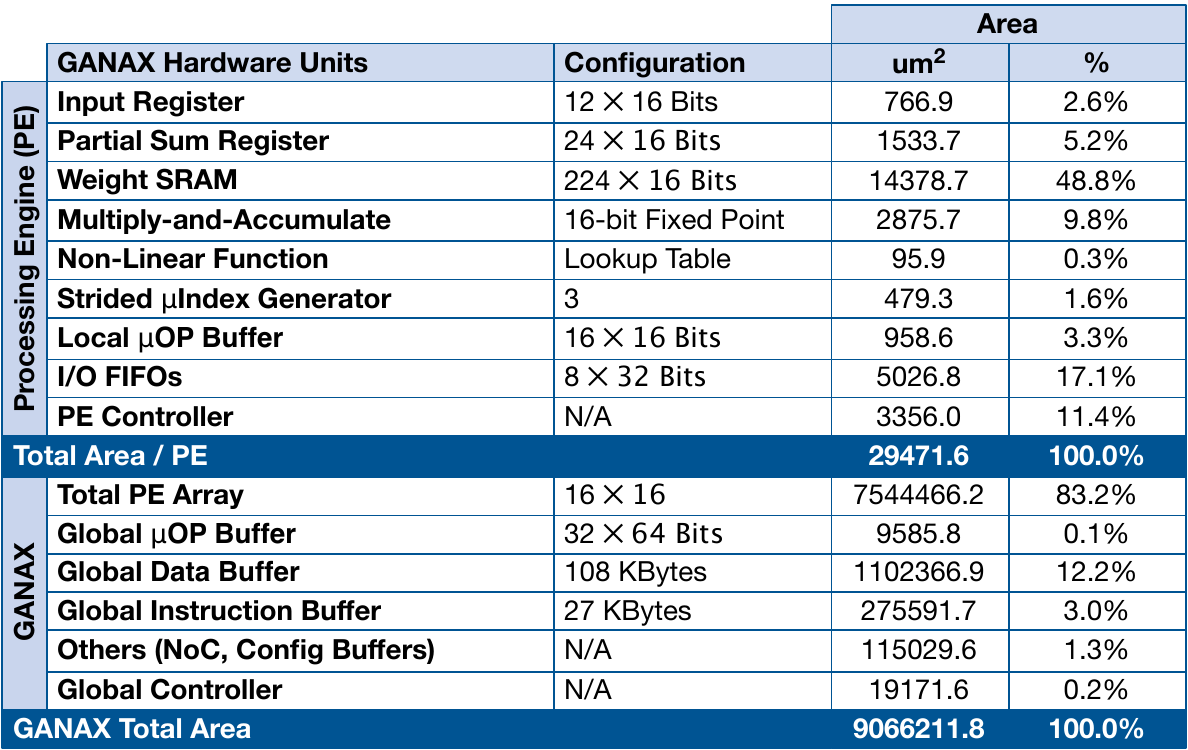}
    \vspace{-0.7cm}
    \label{tab:area}
\end{table}

\section{Evaluation}
\niparagraph{Overall performance and energy consumption comparison.}
Figure~\ref{fig:speedup-eyeriss} depicts the speedup of the generative models with \ganax over \eyeriss~\cite{eyeriss:isca:2016}.
On average, \ganax yields \xx{3.6$\times$} speedup improvement over \eyeriss.
The generative models with a larger fraction of inserted zeros in the input data and larger number of inconsequential operations in transposed convolution layers enjoy a higher speedup with \ganax.
Across all the evaluated models, \bench{3D-GAN} achieves the highest speedup (\xx{6.1}$\times$).
This higher speedup is mainly attributed to its larger number of inserted zeros in its transposed convolution layers. 
On average, the number of inserted zeros for \bench{3D-GAN} is around \xx{80\%} (See Figure~\ref{fig:zero-data}).
On the other extreme, \bench{MAGAN} enjoys a speedup of merely \xx{1.3$\times$}, which is attributed to the lowest number of inserted zeros in its transposed convolution layers compared to other GANs.

\begin{figure}
    \centering
    \subfloat[Runtime]{\includegraphics[width=0.48\textwidth]{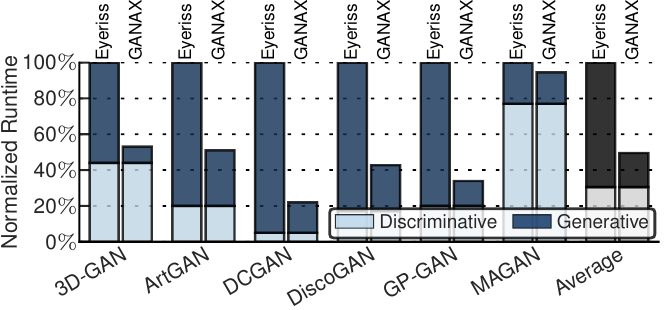}
    \label{fig:runt-bd-eyeriss}}
    \vspace{-0.cm}
    \\
     \subfloat[Energy]{
    \includegraphics[width=0.48\textwidth]{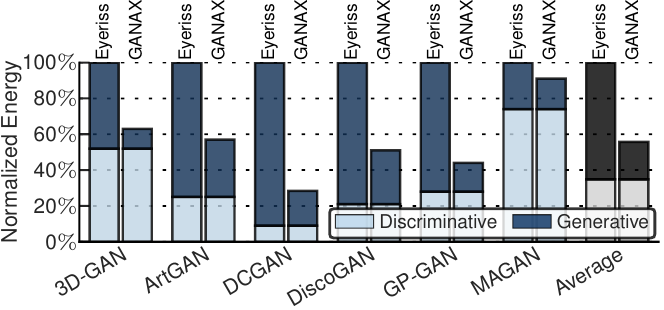}
    \label{fig:enrg-bd-eyeriss}}
    \caption{Breakdown of (a) runtime and (b) energy consumption between discriminative and generative models normalized to the runtime and energy consumption of \eyeriss. For each network, the first (second) bar show the normalized value when the application is executed on \eyeriss (\ganax).}
    \vspace{-0.5cm}
    \label{fig:runt-enrg-bd}
\end{figure}

Figure~\ref{fig:energy-eyeriss} shows the energy reduction achieved by \ganax over \eyeriss.
On average, \ganax effectively reduces the energy consumption by \xx{3.1$\times$} over the \eyeriss accelerator.
The GANs (\bench{3D-GAN}, \bench{DCGAN}, and \bench{GP-GAN}) with the highest fraction of zeros and inconsequential operations in the transposed convolution layers enjoy an energy reduction of more than \xx{4.0$\times$}.
These results reveal that our proposed architecture is efficient in addressing the main sources of inefficiency in the generative models.
Figure~\ref{fig:runt-enrg-bd} shows the normalized runtime and energy breakdown between the discriminative and generative models.   
The first (second) bar shows the normalized runtime (energy) for \eyeriss (\ganax).
To be consistent across all the networks, for the discriminative model of \bench{MAGAN}, we \emph{only} consider the contribution of convolution layers in the overall runtime and energy consumption.  
As the results show, while \ganax significantly reduces both the runtime and energy consumption of generative models, it delivers the same level of efficiency as \eyeriss for the discriminative models. 

\begin{figure}
    \centering
    \includegraphics[width=0.48\textwidth]{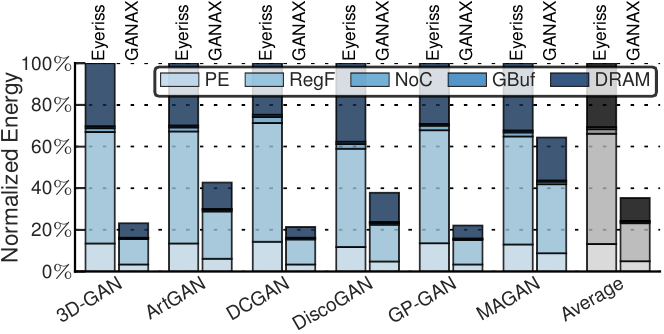}
    \caption{Breakdown of energy consumption of the generative models between different microarchitectural units. The first bar shows the normalized energy breakdown for \eyeriss. The second bar show the energy breakdown for \ganax normalized to \eyeriss.}
    \vspace{-0.2cm}
    \label{fig:enrg_bd_micro}
\end{figure}

\niparagraph{Energy breakdown of the microarchitectural units.}
Figure~\ref{fig:enrg_bd_micro} illustrates the overall normalized energy breakdown of the generative models between distinct microarchitectural components of the \ganax architecture.
The first and second bars show the normalized energy consumed by \eyeriss and \ganax, respectively.
As the results show, \ganax reduces the energy consumption of all the microarchitectural units.
This reduction is mainly attributed to the efficient flow of data in \ganax and the decoupled access-execute architecture that cooperatively diminishes the sources of inefficiency in the execution of transposed convolution operations.

\begin{figure}
    \centering
    \includegraphics[width=0.48\textwidth]{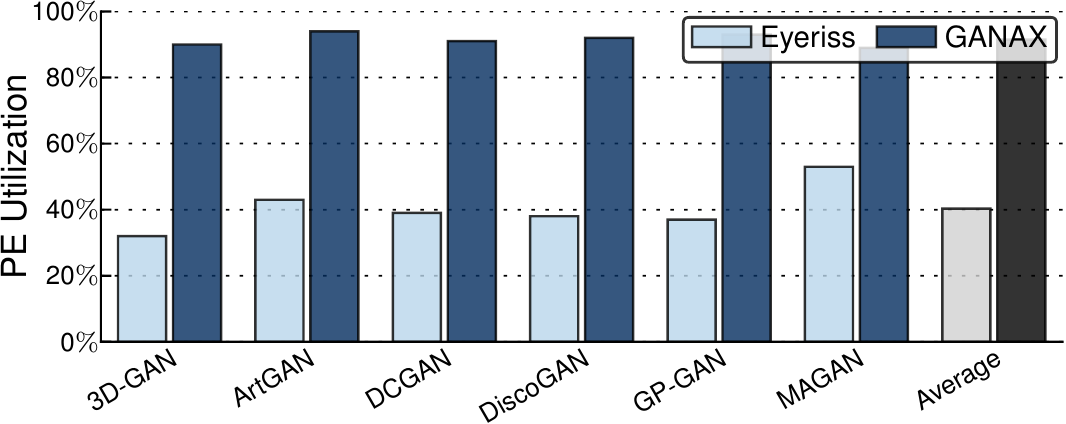}
    \caption{Average PE utilization for the generative models in \eyeriss and \ganax.}
    \vspace{-0.5cm}
    \label{fig:peutil}
\end{figure}

\niparagraph{Processing elements utilization.}
To show the effectiveness of \ganax dataflow in improving the resource utilization, we measure what percentage of the total runtime, the PEs are actively performing a consequential operation.
Figure~\ref{fig:peutil} depicts the utilization of PEs for \eyeriss and \ganax.
\ganax exhibits a high percentage of PE utilization, around \xx{90\%} across all the evaluated GANs.
This high resource utilizations in \ganax is mainly attributed to the proposed dataflow that can effectively force the computation of the rows with similar computation pattern adjacent to each other.
This forced adjacency of similar computation patterns eliminates inconsequential operations, which leads to a significant improvement in the utilization of the processing engines.

\section{Related Work}
\label{sec:related}
\ganax has fundamentally a different accelerator architecture than the prior proposals for deep network acceleration.
In contrast to prior work that mostly focus on convolution operation, \ganax accelerates transposed convolution operation, a fundamentally different operation than conventional convolution.
%
%
Below, we overview the most relevant work to ours along two dimensions: neural network acceleration and MIMD-SIMD acceleration.

\niparagraph{Neural network acceleration.}
Accelerator design for neural networks has become a major line of computer architecture research in recent years.
A handful of prior work explored the design space of neural network acceleration, which can be categorized into ASICs~\cite{bitfusion:isca:2018,tetris:asplos:2017,eyeriss:jssc:2017,scnn:isca:2017,eyeriss:isca:2016,eie:isca:2016,cambricon:micro:2016,cnvlutin:isca:2016,truenorth:arxiv:2016,shidiannao:isca:2015,ngpu:micro:2015,nn:general:pact:2015,diannao:isca:2014,olivier:isca:2013,npu:micro:2012}, FPGA implementations~\cite{tabla:hpca:2016,dnnweaver:micro:2016,snnap:hpca:2015,FPGADeep:fpga:2015,neuflow:cvpr:2011}, using unconventional devices for acceleration~\cite{isaac:isca:2016,prime:isca:2016,anpu:isca:2014}, and dataflow optimizations~\cite{cnn:loop:fpga:2017,pipelayer:hpca:2017,flexflow:hpca:2017,deep-comp:iclr:2016,eyeriss:isca:2016,blocking:systematic:arxiv:2016,divergent:taco:2015}.
Most of these studies have focused on accelerator design and optimization of merely one specific type of convolutional as the most compute-intensive operation in deep convolutional neural networks.

\eyeriss~\cite{eyeriss:isca:2016} proposes a row stationary dataflow that yields high energy efficiency for convolutional operation.
\eyeriss exploits data gating to skip zero inputs and further improves the energy efficiency of the accelerator. However, \eyeriss still wastes cycles for detecting the zero-valued inputs.
Cnvlutin~\cite{cnvlutin:isca:2016} can save compute cycle and energy for zero-values inputs but still wastes resources for zero-valued weights. In contrast, Cambricon-X~\cite{cambricon:micro:2016} can skip zero-valued weights but still wastes compute cycles and energy for zero-input values.
SCNN~\cite{scnn:isca:2017} proposes an accelerator that can skip both zero-valued inputs and weights and efficiently performs convolution on highly sparse data.
However, not only SCNN cannot handle dynamic zero-insertion in input feature maps, but also it is not efficient for non-sparse vector-vector multiplications, which are the dominant operation in discriminative models of GANs.
None of these works can perform zero-insertion into the input feature maps, which is fundamentally a requisite for transposed convolution operation in the generative models.
Compared to these successful prior work in neural network acceleration, \ganax proposes a unified architecture for efficient acceleration of both conventional convolution and transposed convolution operations.
As such, \ganax encompasses the acceleration of a wider range of neural network models.
 
\niparagraph{MIMD-SIMD accelerators.}
While the idea of access-execute is not brand-new, \ganax extends the concept of access-execute architecture~\cite{stream:accl:isca:2017,decoupled:affine:isca:2017,decoupled:prefetching:micro:2016,decoupled:smith:sigarch:1982} to the finest granularity of computation for each individual operand for deep network acceleration.  
A wealth of research has studied the benefits of MIMD-SIMD architecture in accelerating specific applications~\cite{pasm,precision,netra,superb,simd-mimd:image,simd-mimd:reconf,simd-mimd:fpga,mixed-mode:fpga,simd-mimd:augmented}.
Most of these works have focuses on accelerating computer vision applications.
For example, PRECISION~\cite{precision} proposes a reconfigurable hybrid MIMD-SIMD architecture for embedded computer vision.
In the same line of research, a recent work~\cite{simd-mimd:augmented} proposes a multicore architecture for real-time processing of augmented reality applications.
The proposed architecture leverages SIMD and MIMD for data- and task-level parallelism, respectively.      
While these works have studied the benefits of MIMD-SIMD acceleration mostly for computer vision applications, they did not study the potential gains of using MIMD and SIMD accelerators for modern machine learning applications.
Prior to this work, the benefits, limits, and challenges of MIMD-SIMD architectures for modern deep model acceleration was unexplored.
Conclusively, the GANAX architecture is the first to explore this uncharted territory of MIMD-SIMD acceleration for the next generation of deep networks.

\section{Conclusion}
\label{sec:conclusion}
Generative adversarial networks harness both generative and discriminative deep models in a game theoretical framework to generate close-to-real synthetic data.
The generative model uses a fundamentally different mathematical operator, called transposed convolution, as opposed to the conventional convolution operator.
Transposed convolution extrapolates information by first inserting zeros and then applying convolution that needs to cope with irregular placement of none-zero data.
To address the associated challenges for executing generative models without sacrificing accelerator performance for conventional DNNs, this paper devised the \ganax accelerator.
In the proposed accelerator, we introduced a unified architecture that conjoins SIMD and MIMD execution models to maximize the efficiency of the accelerator for both generative and discriminative models. 
On the one hand, to conform to the irregularities in the generative models, which are formed due to the zero-insertion step, \ganax supports selective execution of only the required computations by switching to a MIMD-SIMD mode.
To support this mixed execution mode, \ganax offers a decoupled micro access-execute paradigm at the finest granularity of its processing engines.
On the other hand, for the conventional discriminator DNNs, it sets the architecture in a purely SIMD mode.
The evaluation results across a variety of generative adversarial networks reveal that the \ganax accelerator delivers, on average, 3.6$\times$ speedup and 3.1$\times$ energy reduction for the generative models.
These significant benefits are attained without sacrificing the execution efficiency of the conventional discriminator DNNs.
\section{Acknowledgments}
We thank Hardik Sharma, Ecclesia Morain, Michael Brzozowski, Hajar Falahati, and Philip J. Wolfe for insightful discussions and comments that greatly improved the manuscript.
Amir Yazdanbakhsh is partly supported by a Microsoft Research PhD Fellowship.
This work was in part supported by NSF awards CNS\#1703812, ECCS\#1609823, CCF\#1553192, Air Force Office of Scientific Research (AFOSR) Young Investigator Program (YIP) award \#FA9550-17-1-0274, NSF-1705047, Samsung Electronics, and gifts from Google, Microsoft, Xilinx, and Qualcomm.
%


\bibliographystyle{ieeetr}
\bibliography{paper}

\end{document}